# A novel computation of the linear plasma response to a resonant error field in single-fluid rotating visco-resistive MHD

*Paolo Zanca*

*Consorzio RFX (CNR, ENEA, INFN, Università di Padova, Acciaierie Venete Spa), Padova (Italy)*

**Abstract**

This paper reexamines the linear plasma response to a static resonant error field (EF) in the single-fluid rotating visco-resistive magneto-hydrodinamic (MHD). A tearing-mode stable, rotating plasma shields a resonant static EF by a current sheet at the resonant surface. This response is encapsulated within the delta prime ($\Delta'$), a quantity which measures the magnitude and phase of the current sheet. However, if EF exceeds an amplitude threshold this equilibrium breaks down and a wall-locked tearing mode is formed. Several basic aspects of the problem are addressed. First, we assess the validity of the radial Fourier transform method, commonly used to solve analytically the problem, by comparison with a completely different technique. Second, we derive a new analytical $\Delta'$ global formula valid in a wide range of plasma parameters. This formula describes the $\Delta'$ features much better than previous asymptotic regimes modelling. Third, we derive the EF amplitude threshold for producing a locked mode, pointing out the crucial role of the neoclassical poloidal flow damping effect. The result is almost identical to recent two-fluids outcomes, showing that the choice between single-fluid and two-fluids MHD is not crucial in this specific problem.

## 1. Introduction

Magnetic confinement experiments are inevitably subject to error fields (EF), small amplitude static magnetic fields due to coils and machine imperfections. They represent a concern when they resonate in the plasma, namely when they have the same helicity as the equilibrium field at a given radius (resonant surface). In fact, though in intrinsically tearing-stable rotating plasmas they are shielded by the plasma rotation at the resonant surface, above an amplitude threshold they anyway produce a wall locked magnetic island [1], making the plasma prone to a disruption in tokamaks.

This phenomenon is known as EF penetration. A research line interprets the EF penetration with linear magneto-hydrodynamic (MHD), in both one fluid and two-fluids (drift-MHD) versions [1-6]. The justification for using linear theory is that plasma rotation hinders the EF driven reconnection before the penetration takes place. As pointed out in [1] the EF produces a static "suppressed island", with the plasma flow slipping through it, which can be modelled by linear MHD. The linear plasma response is encapsulated within the delta prime ($\Delta'$), a quantity which measures the magnitude and phase of the induced current sheet, which shields the EF at the resonant surface. The current sheet couples with the radial magnetic field associated to the suppressed island to develop an electromagnetic torque near the resonant surface. This torque is balanced by the viscous torque produced by a modification of the unperturbed velocity profile. However, above an amplitude threshold for the EF this equilibrium breaks down and a non-linear wall-locked magnetic island is formed (EF penetration). Over the years there has been the tendency to enrich the theoretical models, passing from one-fluid theory [1-3, 7] to two fluids drift-MHD [4-6], adding any time new effects. However, in our opinion several basic aspects of the analysis methods have not been adequately discussed. First of all, the applicability of the radial Fourier transform, commonly used to compute $\Delta'$ by analytical means. In fact, a priori the problem does not fulfil the conditions to adopt this technique. Second, the $\Delta'$ analytical formulas are obtained only in asymptotic regimes of the plasma parameters: a question arises on what extent do they effectively describe the $\Delta'$ important features. Third, the most complete model [6] gives a significant variation of predictions of the EF threshold. Some of them look like the experimental scaling laws, others are quite different. For instance, as far as the dependence on the plasma electron density $n_e$ is concerned, they vary from $n_e^0$, which is far from experiments, to $n_e^{0.5}$, which is much closer to them. So, the question arises on which is the crucial physical element determining such a variety of predictions. The present work is focused on these fundamental issues, and to this purpose we consider the minimal model, namely the single-fluid, visco-resistive MHD. The paper is organized as follows. In section 2 the fundamental equations are presented and the tearing mode solution is obtained in the plasma region where ideal MHD is valid (outer region). Section 3 derives the equations in a narrow layer around the resonant surface, where non-ideal terms play a crucial role (inner region). Section 4 presents a new numerical method to solve the equations of the inner region. Section 5 approaches the same problem by numerical solution after the traditional Fourier transformation, and compares the results with those of the previous section. Section 6 derives a new formula for Δ', obtained by an approximate analytical solution of the Fourier transformed equations. The new formula is compared with asymptotic-regimes expressions derived in previous literature. Section 7 estimates the EF penetration threshold, and compares it with published theoretical predictions as well as experimental results. Section 8 draws the conclusions.

## 2. Starting equations

We take the most basic model suitable to our study, namely single-fluid, visco-resistive MHD equations with zero pressure gradient, neglecting the impact of the second-order moments:

1) $\boldsymbol{E} + \boldsymbol{V} \times \boldsymbol{B} = \eta \boldsymbol{J}$

2) $\rho \left( \frac{\partial}{\partial t} + \boldsymbol{V} \cdot \nabla \right) \boldsymbol{V} = \boldsymbol{J} \times \boldsymbol{B} + \mu \nabla^2 \boldsymbol{V}$

3) $\nabla \cdot \boldsymbol{B} = 0. \quad \partial \boldsymbol{B}/\partial t = -\nabla \times \boldsymbol{E}, \quad \mu_0 \boldsymbol{J} = \nabla \times \boldsymbol{B}$.

Equation (1) is Ohm's law with $\eta$ the plasma resistivity. Equation (2) is the motion equation with $\mu$ the perpendicular (dynamic) viscosity and $\rho$ the mass density. All $\eta, \rho, \mu$ are taken spatially constant. Formulas (3) are Maxwell's equations. Since toroidal effects are not expected to be important for the problem under investigation, we describe plasma as a cylindrical configuration with coordinates $(r, \theta, \phi = z/R_0)$ and periodicity length $2\pi R_0$, with $R_0$ the simulated plasma major radius. The plasma minor radius is denoted by $a$. The plasma is surrounded by an ideal shell at $r = b > a$, characterized by the leakage of a single-harmonic error field $b_r(b, \theta, \phi) = b_r^{m,n}(b) \, e^{i(m\theta - n\phi)} + c.c.$, with $m, n$ the poloidal and toroidal mode numbers respectively, and $c.c.$ the complex conjugate term. We split each quantity into the zeroth-order component (depending on $r$ only) and the harmonic component, assumed to be a small perturbation. The non-ideal terms of equations (1), (2), i.e. inertia, viscosity and resistivity, are discarded in most of the plasma. Therefore, (2) simply becomes the force-balance equation $\boldsymbol{J} \times \boldsymbol{B} = \boldsymbol{0}$. By neglecting harmonic coupling terms, the zeroth-order magnetic field $\boldsymbol{B_0} = (0, B_{0\theta}, B_{0\phi})$ is given by the force-free relationship $\mu_0 J_0 = \sigma(r) B_0$, with $\sigma(r) = \mu_0 J_0 \cdot B_0/(B_0 \cdot B_0)$ arbitrary function, alongside Ampere's law $\nabla \times B_0 = \mu_0 J_0$. The perturbed field $b^{m,n}$ is obtained from the linearized force balance equation $J_0 \times b^{m,n} + j^{m,n} \times B_0 = 0$, alongside $\nabla \times b^{m,n} = \mu_0 j^{m,n}$, $\nabla \cdot b^{m,n} = 0$. These equations can be combined to give an ordinary second-order differential equation (Newcomb's equation) for the radial profile of the perturbation $\psi^{m,n}(r,t) = -i\, r b_r^{m,n}(r,t)$ [8]:

$$4)\quad \frac{\partial}{\partial r}\left(\frac{r}{H_{mn}}\frac{\partial}{\partial r}\psi^{m,n}\right) - \left[\frac{1}{r} + \frac{r\, G^{m,n}}{H_{mn} F^{m,n}}\frac{d\sigma}{dr} + \frac{2\, m\, n\, \varepsilon\, \sigma}{H_{mn}^2} - \frac{r\sigma^2}{H_{mn}}\right]\psi^{m,n} = 0$$

$\varepsilon(r) = r/R_0$, $G^{m,n}(r) = mB_{0\phi} + n\varepsilon B_{0\theta}$, $F^{m,n}(r) = mB_{0\theta} - n\varepsilon B_{0\phi}$, $H_{mn}(r) = m^2 + n^2\varepsilon^2$

Equation (4) is singular at the resonant surfaces, identified by the radius $r_{m,n}$ such that $F^{m,n}(r_{m,n}) = 0$. Therefore, it is solved in the separate regions $0 < r < r_{m,n}$ and $r > r_{m,n}$, by imposing the continuity of $\psi^{m,n}$ at $r = r_{m,n}$. The radial derivative of $\psi^{m,n}$ is discontinuous at $r = r_{m,n}$, in general. Since (4) is a second order differential equation, we can represent the solution by the basis $\hat{\psi}_s(r)$, $\hat{\psi}_b(r)$

$$5)\quad \psi^{m,n}(r,t) = \Psi_s(t)\hat{\psi}_s(r) + \Psi_b(t)\hat{\psi}_b(r)$$

where $\hat{\psi}_s(r)$ is the real solution of (4) regular at $r = 0$, with $\hat{\psi}_s(r_{m,n}) = 1$ and $\hat{\psi}_s(b) = 0$, and $\hat{\psi}_b(r)$ is the real solution of (4) with $\hat{\psi}_b(b) = 1$ and $\hat{\psi}_b(r \leq r_{m,n}) = 0$. The complex quantities $\Psi_s, \Psi_b$ define amplitude and phase of the perturbation at $r_{m,n}, b$ respectively. In particular, $\Psi_b$ represents the EF at the shell location, whereas $\Psi_s$ is the reconnected helical magnetic flux at the resonant surface. The delta prime parameter is the normalized radial derivative discontinuity of $\psi^{m,n}$ at the resonant surface:

$$6)\quad \Delta' = \frac{1}{\Psi_s}\frac{d}{dr}\psi^{m,n}\bigg|_{r_{m,n}^-}^{r_{m,n}^+} = \frac{1}{r_{m,n}}\left(E_s + \frac{\Psi_b}{\Psi_s}E_{bs}\right),\quad E_s = r\frac{d}{dr}\hat{\psi}_s\bigg|_{r_{m,n}^-}^{r_{m,n}^+},\quad E_{bs} = r\frac{d}{dr}\hat{\psi}_b\bigg|_{r_{m,n}^-}^{r_{m,n}^+}$$

This quantity represents the amplitude and phase of the helical current sheet flowing at the resonant surface. We assume that the plasma equilibrium is stable to tearing modes in the absence of EF ($\Psi_b = 0$), implying $E_s < 0$ [9]. By denoting $x = (r - r_{m,n})/r_{m,n}$, the solution of (4) has the following expansion for $x \to 0$:

7) $\psi^{m,n}(x) = \Psi_s \left(1 + \lambda_0 x \ln|x| + r_{m,n} \frac{\Delta'}{2}|x| + \cdots \right), \quad \lambda_0 \propto \frac{d\sigma}{dr}(r_{m,n})$

## 3. Singular layer equations

To solve the singularity of (4) at $r = r_{m,n}$ we have to retain all the non-ideal terms of (1, 2) in a layer around the resonant surface, hereafter called 'inner region', to be distinguished from the rest of the plasma which is called 'outer region' (i.e. the region where (4) holds), as it is customary in the traditional literature. It is convenient to normalize equations (1-3) as follows. Lengths are normalized by the resonant radius $r_{m,n}$, magnetic field by the magnetic field strength $B_0$ at the resonant radius, current by $B_0/(\mu_0 r_{m,n})$, velocity by the Alfven velocity $V_A = B_0/\sqrt{\mu_0 \rho}$, time by the Alfven time $\tau_A = r_{m,n}/V_A$, electric field by $B_0 V_A$, resistivity by $\mu_0 r_{m,n} V_A$, viscosity by $\rho r_{m,n} V_A$. The normalized resistivity is $\hat{\eta} = S^{-1}$, with $S = \tau_R/\tau_A$ the Lundquist number (typically $S \gg 1$) and $\tau_R = \mu_0 r_{mn}^2/\eta$ the resistive diffusion time. The normalized viscosity is $\hat{\mu} = S^{-1}P$, with $P = \tau_R/\tau_V$ the Prandl number and $\tau_V = \rho\, r_{mn}^2/\mu$ the viscous diffusion time. Hereafter, all the quantities referring to the inner region are treated as normalized (unless strictly necessary we discard the hat over the quantities for convenience of notation). Since the inner region is assumed narrow, we can adopt here the simplification of slab Cartesian geometry. The slab/cylindrical geometries relationship are the following: $\hat{x}$ corresponds to the radial direction with $x \to (r - r_{m,n})/r_{m,n}$; $\hat{z}$ corresponds to the symmetry direction ($\partial/\partial z = 0$), namely the direction of the zeroth order field at the resonant surface; $\hat{y}$ corresponds to the wave vector direction, namely $yk \to m\theta - n\phi$, with $k$ the normalized wave-number of the perturbation. For the layer quantities, we drop the superscript/subscript $m,n$, by denoting the perturbations with a tilde. We split each quantity into zeroth-order field and perturbation, $\mathcal{A}(x,y) = \mathcal{A}_0(x) + \tilde{\mathcal{A}}(x)e^{iky} + \tilde{\mathcal{A}}^*(x)e^{-iky}$, and we neglect the non-linear coupling. The magnetic field is represented by

8) $\boldsymbol{B} = \nabla\psi \times \hat{\boldsymbol{z}} + B_z\hat{\boldsymbol{z}}$

Note the relationship $k\tilde{\psi} = \psi^{m,n}/(rB_0)$, between $\tilde{\psi}$ and the outer region $\psi^{m,n}$. By using (8), Ampere's law and Faraday's law give respectively

9) $\boldsymbol{J} = -(\nabla^2\psi)\hat{\boldsymbol{z}} + \nabla B_z \times \hat{\boldsymbol{z}}, \qquad E_z = -\partial\psi/\partial t$

We also define the Poisson bracket as $[A,B] = \dfrac{\partial A}{\partial x}\dfrac{\partial B}{\partial y} - \dfrac{\partial A}{\partial y}\dfrac{\partial B}{\partial x} = \nabla A \times \nabla B \cdot \hat{\boldsymbol{z}}$. We take incompressible flow $\nabla \cdot \boldsymbol{V} = 0$, hence

10) $\boldsymbol{V} = \nabla\phi \times \hat{\boldsymbol{z}} + V_z\hat{\boldsymbol{z}}$

with $\phi$ the velocity stream function. Note that $-\nabla \times \boldsymbol{V} \cdot \hat{\boldsymbol{z}} = \nabla^2\phi$, and $\boldsymbol{V} \cdot \nabla A = [A,\phi]$. The $\hat{\boldsymbol{z}}$ component of (1) gives:

11) $\partial\psi/\partial t + [\psi,\phi] = \hat{\eta}\nabla^2\psi$

Before considering the motion equation (2), we note that

12) $\boldsymbol{J} \times \boldsymbol{B} = [B_z,\psi]\hat{\boldsymbol{z}} - (\nabla^2\psi)\nabla\psi - B_z\nabla B_z$

Then, we take $\hat{\boldsymbol{z}} \cdot \nabla \times$ of equation (2) to obtain

13) $\dfrac{\partial}{\partial t}\nabla^2\phi + [\nabla^2\phi,\phi] = [\nabla^2\psi,\psi] + \hat{\mu}\nabla^4\phi$

Equations (11,13) represent a two-field model for the magnetic flux function $\psi$, and the velocity stream function $\phi$. Now, we proceed with their linear analysis. Since the outer region solution for $\psi^{m,n}$ (eq. (4)) is continuous across the resonant surface, i.e. $\psi^{m,n}(r_{s+}) = \psi^{m,n}(r_{s-})$, and the perturbation $\tilde{\psi}$ must connect to the outer solution at the boundaries of the inner region, we take $\tilde{\psi}$ to be an even function of $x$. Since the zeroth-order fields do not depend on $y$, all the Poisson brackets vanish at zeroth-order. Taking into account the second of (9), and the presence of a constant equilibrium inductive electric field $E_{z0}$, the zeroth-order solution of (11) is $\psi_0(x,t) = -x^2/2 - \hat{\eta}t$, global constant apart. Therefore, the whole function $\psi$ is even in $x$: $\psi(-x,y) = \psi(x,y)$. Then, (11, 13) imply that $\phi$ is odd in $x$, $\phi(-x,y) = -\phi(x,y)$, so that we can take $\phi_0(x) = -V_{y0} x$, with $V_{y0}$ constant, as zeroth-order solution of (13). According to (8)-(10), the zeroth-order magnetic field, current and velocity are $\mathbf{B_0} = x\,\hat{\mathbf{y}} + B_{z0}\hat{\mathbf{z}}$, $\mathbf{J_0} = -dB_{z0}/dx\,\hat{\mathbf{y}} + \hat{\mathbf{z}}$, $\mathbf{V_0} = V_{y0}\,\hat{\mathbf{y}} + V_{z0}\hat{\mathbf{z}}$. At the zeroth-order, the $\hat{\mathbf{x}}$ component of (2) gives $\mathbf{J_0} \times \mathbf{B_0} = \mathbf{0}$, implying a force-free equilibrium $\mathbf{J_0} = \sigma \mathbf{B_0}$, with $B_{z0} = \sqrt{B_0^2 - x^2}$ and $\sigma(x) = \mathbf{J_0} \cdot \mathbf{B_0}/(\mathbf{B_0} \cdot \mathbf{B_0}) = (B_0^2 - x^2)^{-1/2}$.

Before considering the perturbations in (11, 13), let us introduce the small parameter $\varepsilon = (\hat{\eta}/k)^{1/3} = (Sk)^{-1/3} \ll 1$. We remember that $\hat{\eta} = S^{-1}$, $\hat{\mu} = S^{-1}P$. The quantity $\varepsilon$ is a normalized length $\varepsilon = \delta/r_{mn}$, with $\delta$ obtained by equating the rate of resistive diffusion to the shear-Alfven frequency: $\eta/(\mu_0 \delta^2) = \delta k/(r_{mn}\tau_A)$.. We will adopt the so-called resistive ordering, which assumes all the terms of the linearized version of (11, 13) to be of the same order in $\varepsilon$. This is realized as follows. Since the perturbation is produced by a static EF over a stable plasma equilibrium, we take $\partial/\partial t = 0$ as customary [4, 6]. Note that this approach is different from the usual linear tearing mode stability analysis, where $\partial/\partial t \neq 0$, but the zeroth-order velocity it taken to be zero ($\mathbf{V_0}=\mathbf{0}$) [9]. Therefore, for the perturbations (11), (13) can be rewritten as

14) $[\widetilde{\psi,\phi}] = \varepsilon^3 k \nabla^2 \tilde{\psi}$

15) $[\widetilde{\nabla^2\phi, \phi}] = [\widetilde{\nabla^2\psi, \psi}] + \varepsilon^3 kP\,\nabla^4 \tilde{\phi}$

Since in the inner region $|x| \ll 1$, we introduce the stretched radial variable $X = \varepsilon^{-1}x$, with $X = O(1)$. The boundaries of the inner region correspond to $X \to \pm\infty$. The zeroth-order fields become $\psi_0(X,t) = -\varepsilon^2 X^2/2 - \hat{\eta}t$, $\phi_0(x) = -\varepsilon V_{y0}X$. To rewrite (14, 15) in terms of $X$, we take into account that $\partial/\partial x = \varepsilon^{-1}\partial/\partial X$, $\partial \mathcal{A}/\partial y = k\tilde{\mathcal{A}}_y(x,y)$ with $\tilde{\mathcal{A}}_y(x,y) = i[\tilde{\mathcal{A}}(x)e^{iky} - \tilde{\mathcal{A}}^*(x)e^{-iky}]$, and $[\mathcal{A}, \mathcal{B}] = k\varepsilon^{-1}(\partial \mathcal{A}/\partial X \, \tilde{\mathcal{B}}_y - \tilde{\mathcal{A}}_y \, \partial \mathcal{B}/\partial X)$. Moreover, for the perturbation we take $\partial^2 \tilde{\mathcal{A}}/\partial X^2$ of the same order of $k^2 \tilde{\mathcal{A}}$ (i.e. strong variation with $x$ in the inner layer), hence $\nabla^2 \mathcal{A} \cong \varepsilon^{-2} \partial^2 \mathcal{A}/\partial X^2$. Equations (14, 15) are then linearized in the perturbations $\tilde{\psi}, \tilde{\phi}$. If we take $V_{y0} = \varepsilon Q$ with $Q = O(1)$, all the linearized terms of (14, 15) are of the same order in $\varepsilon$. Equations (14, 15) finally become

16) $\quad \dfrac{d^2\tilde{\psi}}{dX^2} = iQ\,\tilde{\psi} - iX\,\tilde{\phi}, \qquad Q = (Sk)^{1/3}V_{y0}$

17) $\quad X\dfrac{d^2\tilde{\psi}}{dX^2} = Q\dfrac{d^2\tilde{\phi}}{dX^2} + iP\dfrac{d^4}{dX^4}\tilde{\phi}$

Hereafter, we will assume $Q > 0$.

## 4. Numerical solution of the layer equations

We now discuss the solution of the system (16, 17). It can be equivalently transformed into a sixth order equation for $\tilde{\psi}$, by replacing $\tilde{\phi}$ from (16) into (17), but we do not adopt this solution strategy. The functions $\tilde{\psi}, \tilde{\phi}$ admit the following asymptotic behavior for $X \to \infty$

18) $\quad \tilde{\psi}_\infty(X) = \Psi\left(\dfrac{1}{2}\Delta|X| + 1 + \dfrac{1}{3}Q^2\dfrac{1}{X^2} + o\left(\dfrac{1}{X^4}\right)\right)$

19) $\quad \tilde{\phi}_\infty(X) = \Psi\,Q\left(\dfrac{1}{2}\Delta\,\mathrm{sgn}(X) + \dfrac{1}{X} + \dfrac{1}{3}Q^2\dfrac{1}{X^3} + o\left(\dfrac{1}{X^5}\right)\right)$

with $\Psi$, $\Delta$ arbitrary complex quantities. Relation (18), (19) can be easily verified by insertion into (16), (17). From (18) it follows

20) $\Delta = \frac{1}{\Psi}\left(\frac{d}{dX}\tilde{\psi}\right)_{-\infty}^{+\infty}$

Therefore, $\Delta$ is the normalized delta prime parameter of the inner region. By matching (18) to the outer region expansion (7), and taking into account the adopted inner region normalization, one gets

21) $r_{mn}\Delta' = \varepsilon^{-1}\Delta, \quad \lambda_0 = 0, \quad \Psi_s = r_{mn}B_0\Psi$

The condition $\lambda_0 = 0$ is consistent with the above discussed zeroth-order equilibrium in the inner region having $d\sigma(0)/dX = 0$. Equations (18), (19) represent the *physical* asymptotic behavior that we want to obtain by solving (16), (17). As far as the behavior at $X = 0$ is concerned, from parity requirement we have

22) $\tilde{\phi}(0) = \frac{d^2\tilde{\phi}}{dX^2}(0) = 0, \quad \frac{d}{dX}\tilde{\psi}(0) = 0$

Moreover, by integrating (17) from 0 to $+\infty$, and taking into account of (18), we get

23) $\Psi = \tilde{\psi}(0) + Q\frac{d}{dX}\tilde{\phi}(0) + iP\frac{d^3}{dX^3}\tilde{\phi}(0)$

The solutions of (16), (17) launched from $X = 0$ with arbitrary conditions compatible with (22), exhibit strong exponential divergence for $X \to \infty$ due to the irregular singularity at infinity and the high (six) order of the system. Therefore, it is numerically problematic to combine the independent solutions in order to satisfy the physical trend (18), (19). However, this can be done in a wide range of the parameters $Q, P$ by reducing the order of the problem from six to four. The order reduction follows from introducing the field [10]

24) $\chi = X \frac{d}{dX} \tilde{\psi} - \tilde{\psi}$

Note that $\chi(0) = -\tilde{\psi}(0)$. Moreover, $\chi$ has the same even parity of $\tilde{\psi}$. The functions $\tilde{\psi}, \tilde{\phi}$ can be expressed in terms of $\chi$:

25) $\tilde{\psi} = X \int_0^X \frac{1}{t} \frac{d\chi}{dt} dt - \chi$

26) $\frac{d}{dX} \tilde{\phi} = \mathcal{M}(\chi), \quad \mathcal{M} = i \frac{d}{dX}\left(\frac{1}{X^2} \frac{d}{dX}\right) + Q \frac{1}{X^2}$

Equation (25) is the inverse of (24), whereas (26) stems from the radial derivative of (16). Taking the integral of (17), and making use of (23), (25), (26) one gets a fourth order, non-homogeneous, equation for $\chi$:

28) $iP \frac{d^2}{dX^2} \mathcal{M}(\chi) + Q \mathcal{M}(\chi) - \chi = \Psi$

This is more conveniently written in terms of the variable $z = X^2$ and the normalized field $Y = \chi/\Psi$:

29) $16Pz\frac{d^4}{dz^4}Y + \left(\frac{12P}{z} - \lambda\right)\frac{d^2}{dz^2}Y - \left(\frac{12P}{z^2} - \frac{v}{z}\right)\frac{d}{dz}Y + \left(1 - \frac{\mu}{z^2} - \frac{Q^2}{z}\right)Y + 1 = 0$

$\lambda = 4iQ(P+1), \quad v = 2iQ(3P+1), \quad \mu = 6iQP$

Equation (29) has a regular singularity at $z = 0$ and an irregular singularity for $z \to +\infty$, but less severe than the one of the original system (16), (17). In [10] the second-order equation obtained from (29) with $P = 0$ (i.e. zero viscosity) has been solved analytically with an expansion in terms of Legendre polynomials. This technique does not seem applicable to our fourth order case. Therefore, we rely on the numerical solution. Regularity at $X = 0$ imposes the following two conditions

30) $\frac{d}{dz}Y(0) = -\frac{iQ}{2}Y(0),$

31) $-36P\frac{d^3}{dz^3}Y(0) + 2iQ(P+2)\frac{d^2}{dz^2}Y(0) - \left(\frac{iQ^3}{2} + 1\right)Y(0) = 1$

When considering the homogeneous version of (29) we have to replace 1 by 0 in the right-hand-side of (31). Conditions (30), (31) leave only two independent solutions, $Y_{o1}, Y_{o2}$, of the homogeneous equation, to be added to a particular solution $Y_p$ of the complete equation. The general solution is therefore $Y = \alpha_1 Y_{o1} + \alpha_2 Y_{o2} + Y_p$ with $\alpha_1, \alpha_2$ arbitrary coefficients. From (18), the asymptotic physical behavior of $Y$ for $z \to +\infty$ must be

32) $Y_\infty(z) = -1 - Q^2\frac{1}{z} + o\left(\frac{1}{z^2}\right)$

Therefore, the coefficients $\alpha_1, \alpha_2$ are computed by the least squares solution of the system

33) $\begin{bmatrix} Y_{o1}(\bar{z}) & Y_{o2}(\bar{z}) \\ Y_{o1}'(\bar{z}) & Y_{o2}'(\bar{z}) \\ Y_{o1}''(\bar{z}) & Y_{o2}''(\bar{z}) \end{bmatrix} \begin{bmatrix} \alpha_1 \\ \alpha_2 \end{bmatrix} = \begin{bmatrix} Y_\infty(\bar{z}) - Y_p(\bar{z}) \\ Y_\infty'(\bar{z}) - Y_p'(\bar{z}) \\ Y_\infty''(\bar{z}) - Y_p''(\bar{z}) \end{bmatrix}$

with a suitable choice of the matching point $\bar{z} \gg 1$. In terms of the so obtained $Y$ solution, the parameter (20) is then computed by the following integral

34) $\Delta = 2 \int_0^{+\infty} \frac{1}{\sqrt{z}} \frac{dY}{dz} dz$

We will discuss the results obtained with this technique at the end of the next section, by comparing them to the outcome of a Fourier transform based method.

## 5. Numerical solution of the layer equations by means of the Fourier transform

The Fourier transform method for solving system similar to (16, 17) dates back at least to the work [11]. We adopt the non-unitary definition for the Fourier transform $\bar{\mathcal{A}}(k)$ of a generic function $\mathcal{A}(X)$:

35) $\bar{\mathcal{A}}(k) = \int_{-\infty}^{+\infty} \mathcal{A}(X) e^{-ikX} dX, \qquad \mathcal{A}(X) = \frac{1}{2\pi} \int_{-\infty}^{+\infty} \bar{\mathcal{A}}(k) e^{ikX} dk$

Due to the asymptotic trends (18), (19) the functions $\tilde{\psi}$, $\tilde{\phi}$ are not integrable, but their Fourier transform should exist all the same in the distribution sense. For instance, $\tilde{\psi}$ diverges as $|X|$ for $X \to \infty$, but the Fourier transform of $|X|$ exists ($\propto 1/k^2$ [12]). The possible problem instead concerns the well-known derivative rule

36) $(-ik)^n \bar{\mathcal{A}}(k) = \int_{-\infty}^{+\infty} d^n \mathcal{A}/dX^n \, e^{-ikX} dX$

which holds if

37) $lim_{X \to \infty} d^k \mathcal{A}/dX^k = 0, k = 0, ... n - 1$

In fact, when (37) is satisfied, the terms $d^k \mathcal{A}/dX^k e^{-ikX}\big|_{-\infty}^{+\infty}$, $k = 0, ... n - 1$, coming from integrating by parts $\int_{-\infty}^{+\infty} d^n \mathcal{A}/dX^n \, e^{-ikX} dX$, vanish, so identity (36) is verified. Therefore (37) represents a sufficient condition for (36) [12]. If the rule (36) holds, Fourier transform is convenient to reduce the order of the problem (16, 17), since it converts the derivatives into algebraic factors. In [11] the Fourier transform was applied to interchange modes, whose perturbation profile is localized around the resonant surface. In this case (37) is satisfied and the rule (36) holds. In later works [7, 3, 4, 5, 6] the Fourier transform was applied to the tearing mode layer physics. Nevertheless, according to (18, 19) the fields $\tilde{\psi}$, $\tilde{\phi}$ do not satisfy (37), then there is no guarantee of the validity of (36). No explicit mention of this issue can be found in the above references. The present analysis wants to fill the gap. Let us assume that (36) can be used. Therefore, the Fourier-transformed version of the system (16), (17) is

38) $(k^2 + iQ)\bar{\psi} + \frac{d}{dk}\bar{\phi} = 0$

39) $i\frac{d}{dk}(-k^2\bar{\psi}) + (Q k^2 - iP k^4)\bar{\phi} = 0$

The validity of these two equations will be assessed in the following analysis. The Fourier transform is parity preserving, hence $\bar{\phi}(k)$ is odd and $\bar{\psi}(k)$ is even. The asymptotic behavior of $\bar{\phi}$ for $k \to 0$, consistent with equations (38), (39) is

40) $\bar{\phi}_0(k)/\Psi = -i\pi Q\left[\frac{\Delta}{\pi k} + sgn(k)\right] + \Delta\left(\frac{iQ^3}{2} + 1\right)k + \frac{i\pi Q^3}{6}k^2 sgn(k) + o(k^3)$

This can be verified by direct substitution of (40) into (38), (39). For the arbitrary complex quantities $\Psi, \Delta$ we use the same symbols adopted in (18), (19) for reasons soon evident. The expansion (40) is often reported imprecisely in literature, with the form $\bar{\phi}_0(k) \propto \frac{\Delta}{\pi k} + 1 + o(k)$, which violates the odd parity of $\bar{\phi}(k)$. The form of $\bar{\phi}_0(k)$ mirrors the asymptotic trend (19) for large $X$. Indeed, let us consider the following identity

41) $\bar{\phi}(k) = \int_{-\infty}^{+\infty}\left[\tilde{\phi}(X) - \Psi Q\left(\frac{1}{2}\Delta\, sgn(X) + \frac{1}{X}\right)\right]e^{-ikX}dX - i\pi Q\Psi\left[\Delta\frac{1}{\pi k} + sgn(k)\right]$

We have subtracted and added to $\tilde{\phi}(X)$ the first two terms of (19), and taken into account that the Fourier transform of $sgn(X)$ and $1/X$ are respectively $2/(ik)$ and $-i\pi\, sgn(k)$ [12]. The first integral in the right-hand-side is finite because for $X \to \infty$ the integrand goes to zero as $1/X^3$ (see (19)). Therefore, this integral vanishes as $k \to 0$, since $e^{-ikX} \to 1$ and the integrand becomes an odd parity function. Hence

42) $lim_{k\to 0}\bar{\phi}(k) = lim_{k\to 0}\left\{-i\pi Q\Psi\left[\Delta\frac{1}{\pi k} + sgn(k)\right]\right\}$

In other words, when $k \to 0$ there is consistency between the behavior (42) inferred from (19), and the behavior (40) inferred from equations (38), (39). This is a first positive indication supporting the correctness of equations (38), (39). The decisive proof in this sense comes by comparing the numerical solution of (29) to the numerical solution of the system (38, 39). To this purpose we combine (38, 39) to get the single equation

43) $\frac{d}{dk}\left[\frac{k^2}{k^2+iQ}\frac{d}{dk}\bar{\phi}\right] = k^2(iQ + Pk^2)\bar{\phi}$

A robust method to solve numerically eq (43) involves the Riccati transformation [12]:

44) $w(k) = \frac{k^2}{k^2+iQ}\frac{d\bar{\phi}/dk}{\bar{\phi}} \quad \rightarrow \quad \bar{\phi} \propto e^{\int \frac{k^2+iQ}{k^2} w\, dk}$

Equation (43) is transformed into a first-order non-linear equation for $w$:

45) $\frac{d}{dk}w + \frac{k^2+iQ}{k^2} w^2 = k^2(iQ + Pk^2)$

This method has been discussed in extent in [5]. The convenience of this transformation, from the numerical point of view, will be soon evident. The asymptotic behavior for $k \to 0$ of $w(k)$ is obtained from (40) and (44):

46) $w(k) = \frac{ik}{Q}\left[1 - \frac{\pi}{\Delta} k\, sgn(k)\right] + o(k^2)$

Therefore, the parameter $\Delta$ can be obtained as

47) $\Delta = -2\pi\, lim_{k\to 0+} \frac{dw/dk}{d^2w/dk^2}$

Since (45) is a first order equation, and $w \to ik/Q$ for $k \to 0$, the solution is univocally determined once $P, Q$ are fixed. In a numerical approach, we can integrate backwards (45) up to $k = 0$, starting from a $k_{max}$ large enough, and use (47) to compute $\Delta$. Since the solution is univocally defined, the initial condition $w(k_{max})$ can be *arbitrary*. An illustration of this nice property is given in figure 1. Therefore equation (45) is a very robust technique to compute $\Delta$.

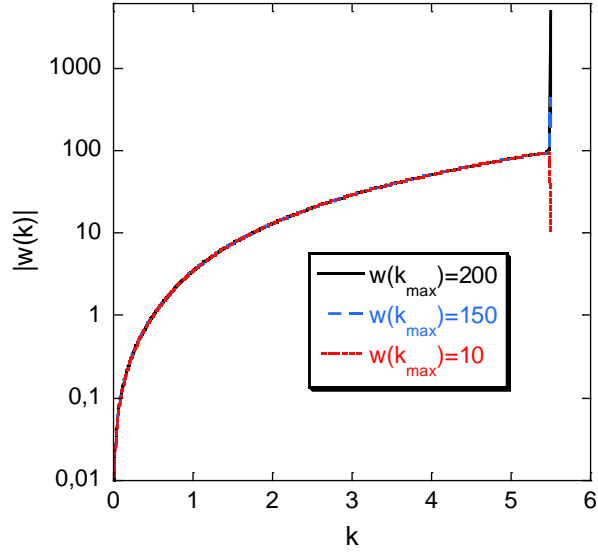

**Fig.1** Absolute value of solutions of (45) obtained with three different initial conditions at $k_{max} = 5.5$. The three solutions merge together very quickly. The y-axis is in log-scale.

Now we compare the $\Delta$ estimates from the numerical methods discussed in this section and in the previous one. In figure 2 we plot the imaginary and real parts of $\Delta(Q, P)$ as computed from solution of equation (29) in combination with relation (34), and from solution of equation (45) in combination with (47). The very good agreement between the two computations is mutually supportive for both the technique adopted. Moreover, it definitely ensures the validity of the Fourier transform method even for the tearing mode layer physics. Note that the dependence on $P$ is weaker than the dependence on $Q$. The imaginary part of $\Delta$ tends to zero as $Q^{-1}$ for $Q \to +\infty$.

Tokamak plasmas exhibit high anomalous perpendicular viscosity, giving rise to a momentum diffusivity similar in magnitude to the energy diffusivity [13]. Instead, plasma resistivity is described by the classical Spitzer formula with neo-classical corrections. Therefore, the physical relevant range of values for the Prandtl number is $P = \tau_R/\tau_v \gg 1$. For example, we estimate $P \sim 10$ for JET low-density discharges ($n_e \sim 10^{19} m^{-3}$, $T_e \sim 2 keV$) [14] and $P \sim 30$ for the ITER inductive scenario considered in [15]. This motivates the choice $P = 11$ in figure 2a.

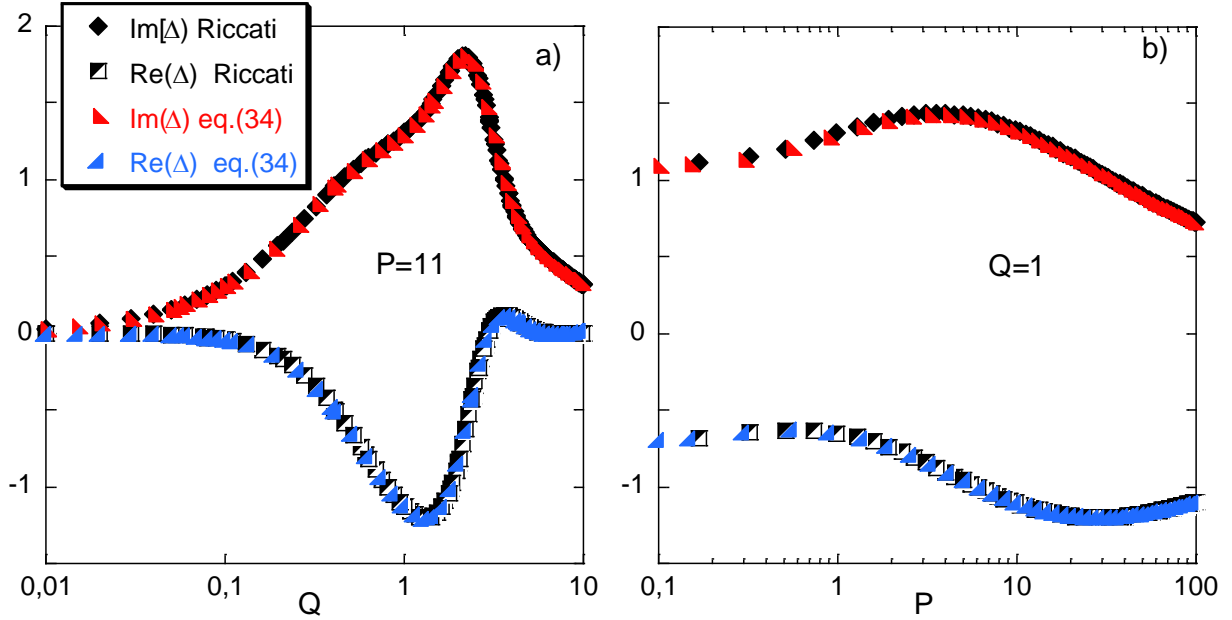

**Fig. 2**. Imaginary and real part of $\Delta$, as a function of $Q$ for $P = 11$ in plot (a), and as function of $P$ for $Q = 1$ in plot (b). The x-axis is in log-scale. In black (diamonds and squares) we report the Fourier-Riccati computation. In red and blue (triangles) we report the computation based on equations (29), (34).

## 6. Approximate analytical solution

In this section we present a novel analytical estimate of $\Delta$, which derives from an approximate solution of (43). First of all, we remove the $1/k$ singularity of $\bar{\phi}$ for $k \to 0$ by introducing the even parity function $\bar{Z}(k) = k\bar{\phi}(k)$. Equation (43) then becomes

48) $(k^2 + iQ)\frac{d^2}{dk^2}\bar{Z} - 2k\frac{d}{dk}\bar{Z} + 2\bar{Z} - [a(k) + b(k) + c(k) + d]\bar{Z} = 0,$

$a(k) = Pk^6, \quad b(k) = iQ(1 + 2P)k^4, \quad c(k) = -Q^2(P + 2)k^2, \quad d = -iQ^3$

Since

49) $d\tilde{\phi}(0)/dX = \frac{i}{2\pi}\int_{-\infty}^{+\infty} \bar{Z}(k)dk$

the integral in the right-hand-side must converge, hence $\bar{Z}(k) \to 0$ for $k \to \infty$. The asymptotic expansion of $\bar{Z}(k)$ for $k \to 0$ stems from (40):

50) $\bar{Z}_0(k) = z_0 + z_1|k| + z_2 k^2 + z_2 k^3 \, sgn(k) + o(k^4)$

$z_0 = -iQ\Psi \Delta, \quad z_1 = -i\pi Q\Psi, \quad z_2 = \Psi\Delta\left(\frac{iQ^3}{2}+1\right), \quad z_3 = \frac{i\pi Q^3}{6}\Psi$

Note that $\Delta$ can be equivalently obtained by both the following relations

51) $\Delta = \pi\frac{z_0}{z_1}, \quad \Delta = -i\pi\frac{Q}{\frac{iQ^3}{2}+1}\frac{z_2}{z_1}$

We didn't find a method to solve the complete equation (48). Instead, we can derive an approximate solution which retains the correct behavior for $k \to 0$, necessary to compute $\Delta$ by (51), and at the same vanishes for $k \to \infty$.

First, we solve an approximation of equation (48) valid for small $k$. If $k^2 < k_1^2 \equiv Q/\sqrt{1+2P}$ both $c(k), d$ are larger (in absolute value) than $a(k), b(k)$, as well as $k^2 < Q$. Therefore, when $k^2 \ll k_1^2$ equation (48) reduces to

52) $iQ\frac{d^2}{dk^2}\bar{Z} - 2k\frac{d}{dk}\bar{Z} + [Q^2(P+2)k^2 + iQ^3 + 2]\bar{Z} = 0, \quad k^2 \ll k_1^2$

To solve it, we get rid of the $k^2 \bar{Z}$ term by the transformation

53) $\bar{Z}(k) = e^{-\Lambda k^2} \bar{U}(k), \quad \Lambda = \frac{i + \sqrt{-1 + iQ^3(2+P)}}{2Q}$

In fact, the ensuing equation for $\bar{U}$ is

54) $\frac{d^2}{dk^2}\bar{U} - \Omega k \frac{d}{dk}\bar{U} + \Xi \bar{U} = 0, \qquad \Omega = 4\Lambda + 2/(iQ), \qquad \Xi = Q^2 - 2\Lambda + 2/(iQ)$

With the variable change $y = \frac{\Omega}{2}k^2$ equation (54) becomes the following generalized hypergeometric differential equation (see section 16.8 of [16]), also known as Kummer's equation (see section 13.2 of [16]),

55) $y \frac{d^2}{dy^2}\bar{U} + \left(\frac{1}{2} - y\right) \frac{d}{dy}\bar{U} + \frac{\Xi}{2\Omega} \bar{U} = 0$

The solution is

56) $\bar{U}(y) = c_1 \, _1F_1\left(-\frac{\Xi}{2\Omega}; \frac{1}{2}; y\right) + c_2 \, y^{1/2} \, _1F_1\left(\frac{1}{2} - \frac{\Xi}{2\Omega}; \frac{3}{2}; y\right),$

with $c_1, c_2$ arbitrary complex coefficients, and $_1F_1$ the generalized hypergeometric function of order $p = q = 1$,

57) $_1F_1(a;b;x) = \sum_{n=0}^{+\infty} \frac{(a)_n}{(b)_n} \frac{x^n}{n!}$,    $(a)_n = a(a+1)\cdots(a+n-1)$

Function $_1F_1$ is also known as confluent hypergeometric function or Kummer's function. Therefore, the solution of (52) is

58) $\bar{Z}(k) = e^{-\Lambda k^2} \left[ c_1 \, _1F_1\left(-\frac{\Xi}{2\Omega}; \frac{1}{2}; \frac{\Omega}{2}k^2\right) + c_2 \, |k| \, _1F_1\left(\frac{1}{2} - \frac{\Xi}{2\Omega}; \frac{3}{2}; \frac{\Omega}{2}k^2\right) \right]$,    $k^2 \ll k_1^2$

Note that the coefficient $c_2$ in (58) is modified with respect to (56). By exploiting (57), the small $k$ expansion of (58) is

59) $\bar{Z}_0(k) = c_1 + c_2 \, |k| - c_1 \frac{\left(\frac{iQ^3}{2}+1\right)}{iQ} k^2 + o(k^3)$,

and comparing (59) with (50), (51) we get

60) $\Delta = \pi \frac{c_1}{c_2}$

The coefficients $c_1, c_2$ can be obtained by imposing at some point $k = k_*$ two constraints for the solution (58), for example for $\bar{Z}$ and its first derivative. Schematically, we write

61) $\bar{Z}(k_*) = G, \ \bar{Z}'(k_*) = G'$

with $k_*, G, G'$ generic. We will discuss later on how choosing $k_*, G, G'$. Then, by combination of (58), (60), (61) we get the following general expression for $\Delta$:

62) $\Delta = \mathcal{D}(Q, P; k_*, G, G') = \pi \left\{ 3 \,_1F_1\left(\frac{3\zeta(Q,P)+\vartheta(Q)}{4\zeta(Q,P)}; \frac{3}{2}; \frac{k_*^2}{Q}\zeta(Q,P)\right) \left\{ k_* G' - G\left[1 - \frac{k_*^2}{Q}(i + \zeta(Q,P))\right]\right\} - G\frac{k_*^2}{Q}[3\zeta(Q,P) + \vartheta(Q)] \,_1F_1\left(\frac{7\zeta(Q,P)+\vartheta(Q)}{4\zeta(Q,P)}; \frac{5}{2}; \frac{k_*^2}{Q}\zeta(Q,P)\right)\right\} \times$

$\left\{ 3 \,_1F_1\left(\frac{\zeta(Q,P)+\vartheta(Q)}{4\zeta(Q,P)}; \frac{1}{2}; \frac{k_*^2}{Q}\zeta(Q,P)\right)\left[-G\frac{k_*}{Q}(i + \zeta(Q,P)) - G'\right] + 3G\frac{k_*}{Q}[\zeta(Q,P) + \vartheta(Q)] \,_1F_1\left(\frac{5\zeta(Q,P)+\vartheta(Q)}{4\zeta(Q,P)}; \frac{3}{2}; \frac{k_*^2}{Q}\zeta(Q,P)\right)\right\}^{-1}$,

with

63) $\zeta(Q, P) = \sqrt{-1 + iQ^3(2 + P)}, \quad \vartheta(Q) = 3i - Q^3$

An important result is that the choice of $k_*, G, G'$ *is not crucial* in (62). By taking $k_* = \sqrt{Q}$ (this choice seems natural looking at (62), and will be better justified later on) we define $\Delta_1(Q, P) = \mathcal{D}(Q, P; Q^{1/2}, 1, 0)$ and $\Delta_2(Q, P) = \mathcal{D}(Q, P; Q^{1/2}, 0, 1)$: $\Delta_1$ refers to the case $|G'/G| \ll 1$, whereas $\Delta_2$ refers to the opposite case $|G'/G| \gg 1$. As shown in figure 3, both $\Delta_1$ and $\Delta_2$ are rather close to the numerical $\Delta(Q, P)$ as obtained in the previous paragraphs (the Fourier-Riccati computation is considered). In other words, the function $\Delta(Q, P)$ is almost completely embedded into (62), and a suitable choice of $k_*, G, G'$ *is merely a refinement*. Therefore, we can take $k_*, G, G'$ according to any sensible and simple argument, for example by connecting the solution (58) to the solution of (48) which is well behaved for $k \to \infty$. To determine this solution, we note that in (48) $a(k)$ is larger (in absolute value) than $b(k), c(k), d$, as well as $k^2 > Q$, when $k^2 > k_2^2 \equiv 2Q + Q/P$. Note that $k_2 > \sqrt{Q} > k_1$.

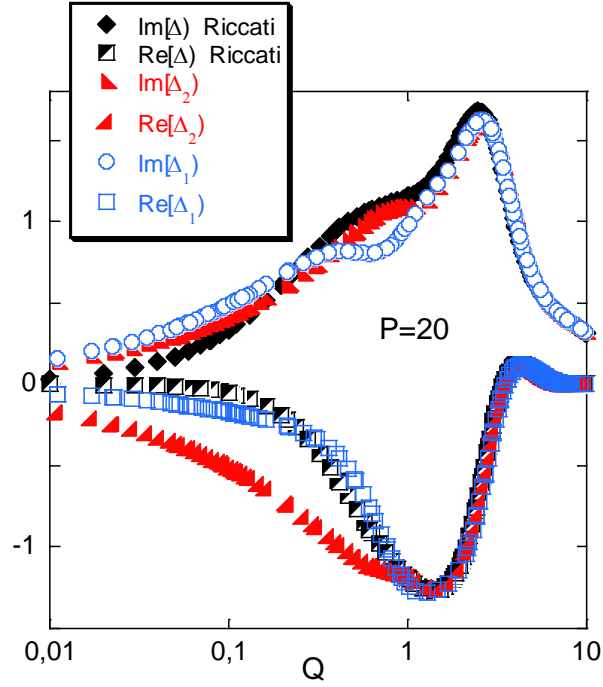

**Fig.3** Imaginary and real part of $\Delta(Q,P)$ as computed by the Fourier-Riccati method (black diamonds and squares), alongside the same quantities for $\Delta_1(Q,P)$ (blue empty circles and squares) and $\Delta_2(Q,P)$ (red triangles), as function of $Q$ for $P=20$. Note the log scale for the x-axis.

Therefore, to study the behavior of $\bar{Z}$ for large $k$ we consider the reduced equation valid for $k^2 \gg k_2^2$

64) $\quad k^2 \frac{d^2}{dk^2}\bar{Z} - 2k\frac{d}{dk}\bar{Z} + 2\bar{Z} - Pk^6\bar{Z} = 0, \qquad k^2 \gg k_2^2$

With the variable change $y = \sqrt{P}k^3/3$ equation (64) becomes

65) $\quad 9y^2\frac{d^2}{dy^2}\bar{Z} + 2\bar{Z} - 9y^2\bar{Z} = 0$

With the further transformation $\bar{Z} = \sqrt{y}\, \bar{U}$, we get the modified Bessel equation

66) $y^2 \frac{d^2}{dy^2} \bar{U} + y \frac{d}{dy} \bar{U} - \left(\frac{1}{36} + y^2\right) \bar{U} = 0$

whose solution, going to zero for $y \to \infty$ is $\bar{U} = K_{\frac{1}{6}}(y)$. Therefore, the solution of (64) which vanishes for $k \to \infty$ is

67) $\bar{Z}_\infty(k) = \left(\frac{\sqrt{P}k^3}{3}\right)^{1/2} K_{\frac{1}{6}}\left(\frac{\sqrt{P}k^3}{3}\right), \quad k^2 \gg k_2^2$

Then, we match solution (58), and its first derivative, to solution (67) at a point $k_*$ chosen between $k_1$ and $k_2$. Hence, we take $k_* = \sqrt{Q}$ as said above. The constraints (61) for the solution (58) are

$G = \bar{Z}_\infty(\sqrt{Q}) = \left(\frac{\sqrt{PQ^3}}{3}\right)^{1/2} K_{\frac{1}{6}}\left(\frac{\sqrt{PQ^3}}{3}\right)$ and $G' = \left.\frac{d\bar{Z}_\infty}{dk}\right|_{k=\sqrt{Q}} = \left.\frac{d}{dk}\left[\left(\frac{\sqrt{P}k^3}{3}\right)^{1/2} K_{\frac{1}{6}}\left(\frac{\sqrt{P}k^3}{3}\right)\right]\right|_{k=\sqrt{Q}}$.

The procedure is not rigorous, but it is justified by the little relevance of the parameters of $k_*, G, G'$ in formula (62), as pointed out before. With these assumptions, formula (62) provides our analytic estimate $\Delta_{an}(Q,P)$:

68) $\Delta_{an}(Q,P) = \mathcal{D}\left[Q,P; \sqrt{Q}, \left(\frac{\sqrt{PQ^3}}{3}\right)^{1/2} K_{\frac{1}{6}}\left(\frac{\sqrt{PQ^3}}{3}\right), \left.\frac{d}{dk}\left[\left(\frac{\sqrt{P}k^3}{3}\right)^{1/2} K_{\frac{1}{6}}\left(\frac{\sqrt{P}k^3}{3}\right)\right]\right|_{k=\sqrt{Q}}\right]$

with $\mathcal{D}$ defined in (62). Indeed, $\Delta_{an}(Q,P)$ follows fairly well the numerical $\Delta(Q,P)$: see figure 4 (red, blue symbols) and figure 5 (grey symbols). $\Delta_{an}(Q,P)$ is somehow less accurate for $P \lesssim 3$ in figure 4b) and for $P \lesssim 0.3$ in figure 5b), but these are regions of little relevance for the present experiments where $P \gg 1$.

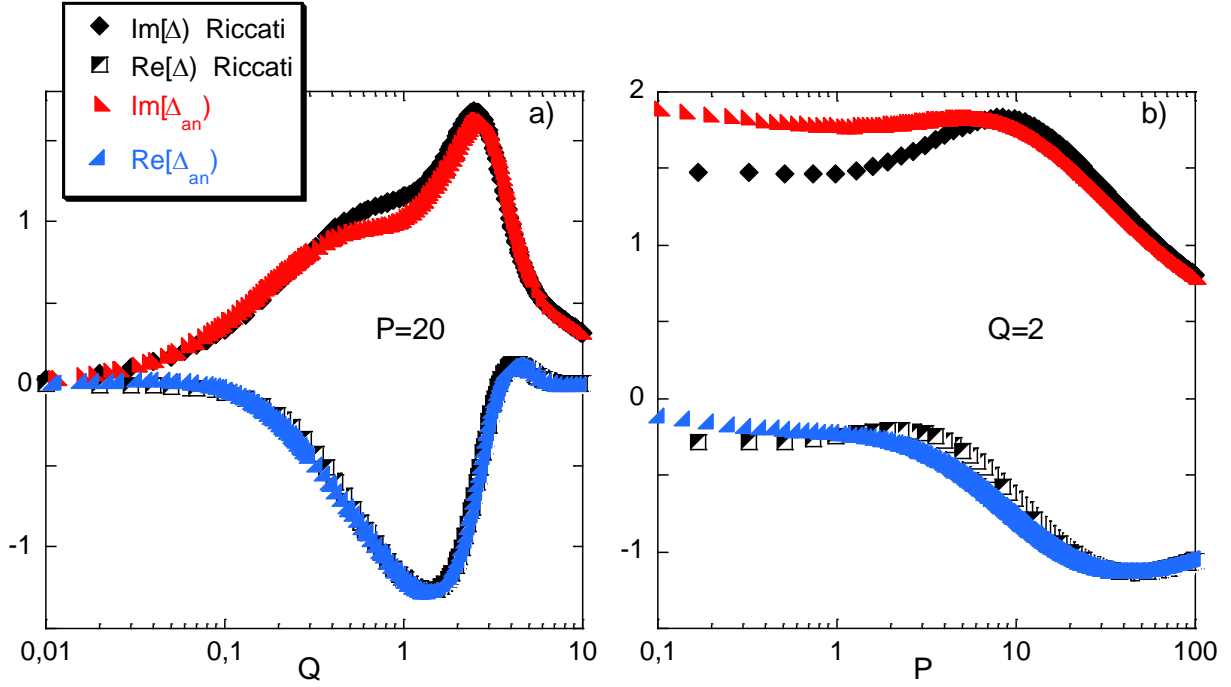

**Fig. 4**. Imaginary and real part of Δ, as function of $Q$ for $P = 20$ in plot (a), and as function of $P$ for $Q = 2$ in plot (b). The x-axis is in log-scale. In black (diamonds and squares) we report the Fourier-Riccati computation. In red and blue (triangles) we report the analytic result (68).

## 7. Comparison with previous analytical results

Now we compare the present theory with well-known analytical predictions obtained in the single-fluid visco-resistive MHD [2]. In figure 1 of [2] four different regimes in the $Q, P$ parameter space are reported: the visco-resistive regime (VR) valid for $Q < 1, Q^{3/2} < P < Q^{-3}$, the visco-inertial regime (VI) valid for $P > 1, P^{-1/3} < Q < P^{1/3}$, the resistive-inertial regime (RI) valid for $Q < 1, P < Q^{3/2}$, the inertial regime (I) valid for $Q > 1, P < Q^3$. Note that the (RI) is not relevant for the present experiments, but it is included for completeness. According to [2], these regimes are expressed by simple power-law formulas for Δ:

69) $\Delta_{VR} = 2.104\, i\, Q\, P^{1/6}, \quad \Delta_{VI} = -4.67\, e^{-i\pi/8}\, Q^{-1/4}\, P^{-1/4}, \quad \Delta_{RI} = -2.124\, e^{-i3\pi/8}\, Q^{5/4},$
$\Delta_I = 3.142\, i\, Q^{-1}$

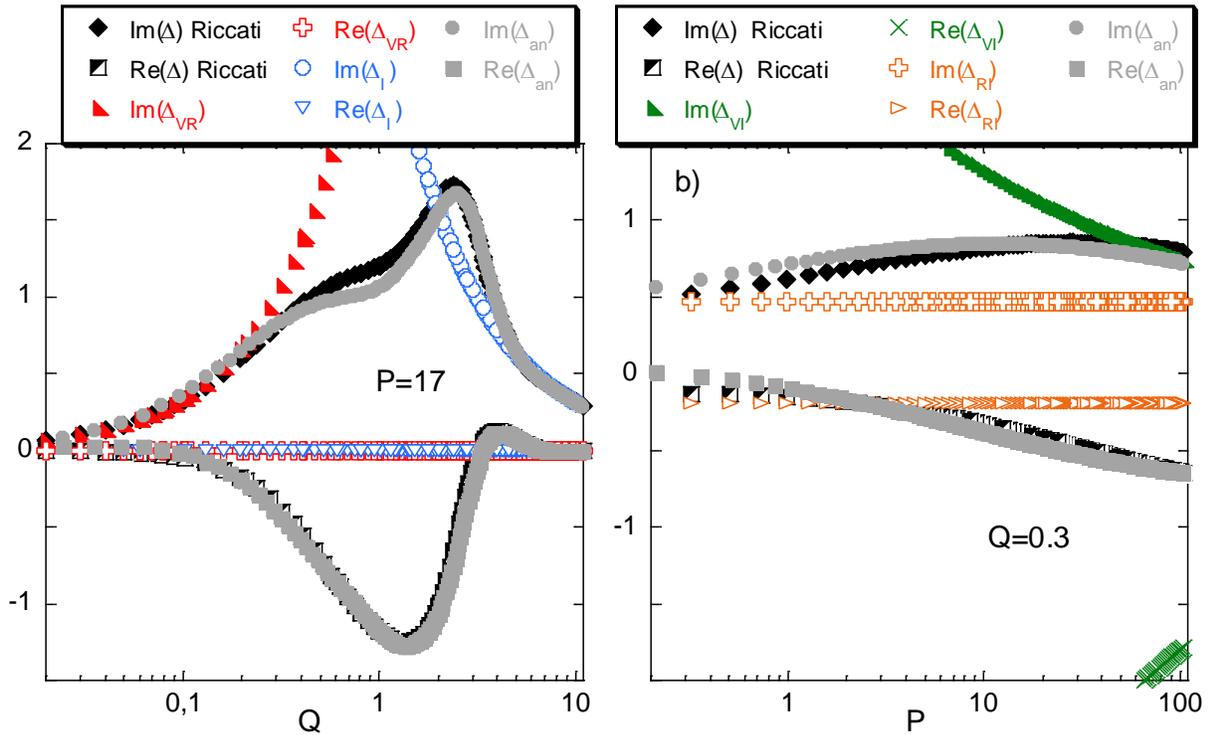

**Fig. 5**. Imaginary and real part of $\Delta$, as function of $Q$ for $P = 17$ in plot (a), and as function of $P$ for $Q = 0.3$ in plot (b). The x-axis is in log-scale. In black (diamonds and squares) we report the Fourier-Riccati computation and in grey (circles and squares) our analytical result (68). Colored symbols refer to the analytical formulas (69). In plot a) we display $\Delta_{VR}$ in red (triangles and crosses) and $\Delta_I$ in blue (empty circles and triangles). In plot b) we display $\Delta_{VI}$ in green (triangles and x) and $\Delta_{RI}$ in orange (empty crosses and triangles)

Expressions (69) differ from those reported in [2] by the factor $P^{1/6}$, as a matter of a different normalization for $\Delta$ there adopted. Formulas (69) are obtained by approximate solutions of equation (43), considered in different intervals of the parameters $Q, P$. The solution method is described in Appendix B of [7], and it is different from the procedure here presented in section 6. The comparison between the prediction of (69) and our results, both numerical and analytical, is displayed in figure 5. Let us consider plot a) where $P = 17$. According to figure 1 of [2] $\Delta_{VR}$ applies for $Q < 17^{-1/3} \approx 0.4$: in this interval $Im(\Delta_{VR})$ agrees pretty well with our result, but $Re(\Delta_{VR}) = 0$ does not, being fixed at zero. Likewise, $\Delta_I$ applies for $Q > 17^{1/3} \approx 2.6$: in this interval $Im(\Delta_I)$ agrees with our result, whereas $Re(\Delta_I) = 0$ is slightly different from it, being fixed at zero. Now we consider plot b) where $Q = 0.3$. $\Delta_{RI}$ applies for $P < 0.3^{3/2} \approx 0.16$: in this interval both $Im(\Delta_{RI})$ and $Re(\Delta_{RI})$ agrees fairly well with our result. Likewise, $\Delta_{VI}$ applies for $P > 0.3^{-3} \approx 37$: in this interval $Im(\Delta_{VI})$ agrees very well with our result, whereas $Re(\Delta_{VI})$ is very far from it.

In conclusion, the agreement between (69) and the asymptotic behavior of our prediction is quite good for the imaginary part of $\Delta$, but for the real part is only partial. Moreover, expressions (69) are too simple to provide an adequate overall description of the function $\Delta(Q, P)$, and in particular they miss the most important feature of the $Q$ dependence, namely the maximum of the imaginary part and the minimum of the real part. Mostly important, the simplification (69) is not necessary to extract predictions from $\Delta(Q, P)$. In the next section we will derive in a very simple way the EF penetration threshold from both our numerical and analytical $\Delta(Q, P)$ estimate.

## 8. Error field penetration threshold

At the resonant surface a static EF induces both a current sheet, mainly as a consequence of the plasma rotation (in the plasma frame of reference EF is seen rotating), and a non-zero radial field, as a consequence of the partial reconnection (partial because it is hindered by the current sheet). The coupling between these two effects gives rise to an electromagnetic torque $T_{EM}$ in the vicinity of the resonant surface. In stationary condition $T_{EM}$ is balanced by the viscous torque produced by the velocity profile modifications. However, above a threshold value for the EF amplitude this balance breaks down, giving rise to a wall-locked island [1]. Now we express this condition by adopting the formalism described in [1]. First of all, $T_{EM}$ can be expressed in a relatively simple way by the quasi-linear physics [1, 8, 17]:

70) $T_{EM,\phi} = \frac{8\pi^2 R_0}{\mu_0} \frac{n\, r_{mn}}{H_{mn}} |\Psi_s|^2 Im(\Delta'), \quad T_{EM,\theta} = -m/n\, T_{EM,\phi}$

The toroidal and poloidal component of $T_{EM}$ given in (70) are intended to be integrated in the angular coordinates and radially integrated in the vicinity of the resonant surface. Formula (70) can be understood in this way: $i\Psi_s$ gives the radial field at the resonant surface, $\Psi_s \Delta'$ the current sheet, so that the real part of the product $\Psi_s \Delta' (i\Psi_s)^*$ gives the angular integrated electromagnetic torque. Note that $T_{EM}$ is a second-order effect in the motion equation (2). Nonetheless, it produces important modifications in the velocity profiles. To this purpose it is convenient writing (2) in terms of the flux surface averaged toroidal and poloidal angular velocities $\Omega_\phi = V_\phi/R_0$, $\Omega_\theta = V_\theta/r$. More precisely, we write (2) for the variation $\Delta\Omega_\phi = \Omega_\phi - \Omega_{\phi 0}$, $\Delta\Omega_\theta = \Omega_\theta - \Omega_{\theta 0}$ of the angular velocities with respect their unperturbed values (denoted by the subscript '0'), i.e. the values in the absence of EF and consequently of $T_{EM}$:

71) $\rho \frac{\partial}{\partial t} \Delta\Omega_\phi = \frac{1}{r}\frac{\partial}{\partial r}\left(\mu r \frac{\partial}{\partial r}\Delta\Omega_\phi\right) + \frac{1}{4\pi^2 r R_0^3} T_{EM,\phi}\, \delta(r - r_{mn})$

72) $\rho \frac{\partial}{\partial t} \Delta\Omega_\theta = \frac{1}{r^3}\frac{\partial}{\partial r}\left(\mu r^3 \frac{\partial}{\partial r}\Delta\Omega_\theta\right) - \frac{\rho}{\tau_\theta}\Delta\Omega_\theta + \frac{1}{4\pi^2 r^3 R_0} T_{EM,\theta}\, \delta(r - r_{mn})$

The second term in the r.h.s of (72) is ad-hoc, and represents the neoclassical poloidal flow damping by means of a characteristic time $\tau_\theta$. We adopt the expression given in [15]: $\tau_\theta \propto a^2/R_0^2\, \tau_i$, with $\tau_i \propto T_e/n_e$ the ion collision time, and $T_e$, $n_e$ the electron temperature and particle density respectively. This way of representing the poloidal flow damping is phenomenological, but it is often used [15, 17]. The Dirac delta formalizes the local character of $T_{EM}$ near the resonant surface. The EF frequency, seen in the frame rotating with the plasma at the resonant surface is $\omega = m\Omega_\theta(r_{mn}) - n\Omega_\phi(r_{mn}) = \boldsymbol{k}\cdot\boldsymbol{V}$, with $\boldsymbol{k} = m/r_{mn}\hat{\boldsymbol{e}}_\theta - n/R_0\hat{\boldsymbol{e}}_\phi$ the wave vector of the perturbation. By imposing the boundary conditions $\Delta\Omega_\phi'(0) = \Delta\Omega_\theta'(0) = \Delta\Omega_\phi(a) = \Delta\Omega_\theta(a) = 0$, the continuity at $r = r_{mn}$, and assuming equilibrium ($\partial/\partial t = 0$), (71), (72) can be combined to give the following torque-balance equation [17]

73) $\frac{1}{2\tau_A^2 r_{m,n} H_{mn}(r_{mn})} \left(\frac{|\Psi_s|}{B_0}\right)^2 Im(\Delta') = (\tau_V \mathcal{J})^{-1}(\omega_0 - \omega)$, $\quad \mathcal{J} = \left[n^2 \varepsilon(r_{m,n})^2 \ln\left(\frac{a}{r_{m,n}}\right) + m^2 F\left(\sqrt{\frac{\tau_V}{\tau_\theta}}\right)\right]$

where $\omega_0$ denotes the unperturbed value of $\omega$ and $F$ is the following positive function

74) $F(x) = \frac{I_1(x r_{m,n}/a)[K_1(x r_{m,n}/a)I_1(x) - I_1(x r_{m,n}/a)K_1(x)]}{I_1(x)}$

Experimental estimate gives $\tau_V/\tau_\theta \sim 10^3$, hence we assume $\tau_V/\tau_\theta \gg 1$. In this case we can exploit the useful approximation $F(x) \cong a/(2 r_{m,n} x)$ valid for $x \gg 1$. The left-hand-side of (73) represents a combination of toroidal and poloidal electromagnetic torques. The right-hand-side is the viscous torque produced by velocity profile modifications ($\omega \neq \omega_0$). In $\mathcal{J}$, the term $n^2 \varepsilon(r_{m,n})^2 \ln\left(\frac{a}{r_{m,n}}\right)$ is related to the toroidal flow, the term $m^2 F(\sqrt{\tau_V/\tau_\theta}) \cong m^2 a/(2 r_{m,n})(\tau_V/\tau_\theta)^{-0.5}$ to the poloidal flow. For the typical case of EF resonating in the edge plasma region ($m > n$) the second term should be larger than the first. For instance, let us take $m = 2, n = 1$, $a/r_{m,n} = 0.8$, $a/R_0 = 0.3$, $\tau_V/\tau_\theta = 10^3$: we get $n^2 \varepsilon(r_{m,n})^2 \ln\left(\frac{a}{r_{m,n}}\right) \approx 0.013$ and $m^2 F(\sqrt{\tau_V/\tau_\theta}) \approx 0.08$. It turns out that we can adopt the simplification $\mathcal{J} \propto \left(\frac{\tau_V}{\tau_\theta}\right)^{-\vartheta}$ with $\vartheta \lesssim 0.5$. Indeed, $\vartheta \approx 0.4$ is a good approximation for a $m = 2, n = 1$ mode as long as $50 \lesssim \tau_V/\tau_D \lesssim 5 \cdot 10^3$.

By exploiting (6) we can express $\Psi_s$ in terms of $\Psi_b$ in (73):

75) $\frac{E_{bs}^2}{2\tau_A^2 r_{mn} H_{mn}(r_{mn})} \left(\frac{|\Psi_b|}{B_0}\right)^2 \frac{Im(\Delta')}{|r_{m,n}\Delta' - E_s|^2} = (\tau_V \mathcal{J})^{-1}(\omega_0 - \omega)$

Before EF penetration, the rotation of the plasma in the vicinity of the rational surface is fast ($\omega \sim \omega_0$) so the EF induced reconnection is suppressed. Therefore $|\Psi_s/\Psi_b| \ll 1$, and from (6) $r_{m,n}\Delta' \gg E_s$, so we neglect $E_s$ in (75), as customary [2, 4]. We also replace $\Delta'$ with the inner-region normalized $\Delta$, by the first of (21):

$$76) \quad \frac{b^2 E_{bs}^2}{2\,\tau_A^2\, r_{mn}^2 H_{mn}(r_{mn})} \left[\frac{|b_r(b)|}{B_0}\right]^2 (Sk)^{-1/3} \frac{Im(\Delta)}{|\Delta|^2} = (\tau_V \mathcal{J})^{-1}(\omega_0 - \omega), \quad |b_r(b)| = |\Psi_b|/b$$

According to the previous analysis $Im(\Delta) > 0$, hence $\omega < \omega_0$, namely the EF slows down the plasma rotation at the resonant surface. Note the cylindrical/slab relationship $\omega = \mathbf{k} \cdot \mathbf{V} \to k/\tau_A\, V_{y0}$, with $k$, $V_{y0} = \varepsilon Q$ the normalized wave number and velocity of the inner region, defined in section 3. We exploit this relationship to replace $\omega$ and $\omega_0$ in (76) by $Q = k^{-2/3} S^{1/3} \tau_A \omega$ and $Q_0 = k^{-2/3} S^{1/3} \tau_A \omega_0$. Finally, by using the approximation $\mathcal{J} \propto \left(\frac{\tau_V}{\tau_\theta}\right)^{-\vartheta}$ and discarding inessential geometrical factors, we rewrite (76) as

$$77) \quad \left(\frac{|b_r(b)|}{B_0}\right)^2 \propto \tau_A \tau_V^{\vartheta-1} \tau_\theta^{-\vartheta} \mathcal{T}, \quad \mathcal{T}(Q_0, Q, P) = \Upsilon(Q, P)(Q_0 - Q), \quad \Upsilon(Q, P) = \frac{|\Delta(Q,P)|^2}{Im[\Delta(Q,P)]}$$

$\mathcal{T}$ is a positive quantity for $Q < Q_0$. Moreover it vanishes for $Q = Q_0$ and for $Q = 0$ (due to $\Delta(0, P) = 0$). Therefore, it has a maximum for $0 < Q < Q_0$. We call $m(Q_0, P) = max_Q[\mathcal{T}(Q_0, Q, P)]$. Due to this maximum in $Q$, the right-hand-side of (77) is upper limited. When the normalized EF amplitude (the left-hand-side) exceeds this limit, the torque-balance relation (76) is no more satisfied and this linear analysis breaks down. Indeed, a new non-linear regime associated to a wall-locked magnetic island takes place (EF penetration). Therefore, the EF penetration threshold scales as

78) $\left(\frac{|b_r(b)|}{B_0}\right)_{thr} \propto \tau_A^{1/2} \tau_V^{(\vartheta-1)/2} \tau_\theta^{-\vartheta/2} \, m(Q_0, P)^{1/2}$

Let us assume that we can use the following power-law approximation for $m$: $m(Q_0, P) \propto Q_0^{1+\beta} P^\gamma$. This simplification will be justified later. Then, we develop (78) by taking $\tau_V \propto \tau_E$, with $\tau_E$ the energy confinement time, $\tau_\theta \propto \frac{a^2}{R_0^2} \frac{T_e}{n_e}$ [15], $Q_0 \propto S^{1/3} \tau_A \omega_0$, Spitzer resistivity to compute $\tau_R$, and $\omega_0$ of the order of diamagnetic frequency as customary [1, 4, 6]: $\omega_0 \propto \frac{T_e}{a^2 B_0}$. We discard terms related to the $q$ profile. The temperature is expressed by exploiting the $\tau_E$ definition

79) $\tau_E \propto a^2 R_0 \frac{n_e T_e}{P}, \quad P = V_\phi I_p \wp$

with $P$ the input power, $V_\phi$ the loop voltage, $\wp$ the ratio between total and ohmic power. Moreover, $I_p \propto a^2 B_0 / R_0$, and $V_\phi$ is given by the on-axis Ohm's law with Spitzer resistivity

80) $V_\phi \propto R_0 \eta(0) J_\phi(0) \propto T_e^{-3/2} B_0$

By considering the neo-alcator scaling $\tau_E \propto n_e a R_0^2$ [19], suitable for low-density ohmic ($\wp = 1$) tokamak, we finally get

81) $\left(\frac{|b_r(b)|}{B_0}\right)_{thr,Alc} \propto n_e^{\vartheta + \frac{1}{12}(2\beta - 6\gamma - 1)} B_0^{-\frac{11}{15} - \frac{1}{30}[7\beta + 18(\vartheta - \gamma)]} a^{-\frac{1}{30}(1 + \beta - 24\gamma + 24\vartheta)} R_0^{2\vartheta - 1 - \gamma}$

If we take the ITER89-P scaling $\tau_E \propto I_p^{0.85} B_0^{0.2} P^{-0.5} n_e^{0.1} a^{0.3} R_0^{1.2}$ [19] suitable for L/H-mode tokamak we get

82) $$\left(\frac{|b_r(b)|}{B_0}\right)_{thr,89P} \propto n_e^{0.17+0.74\vartheta-0.22\beta-0.24\gamma} B_0^{-1.03+0.22\beta+0.3\gamma-0.3\vartheta} \times$$
$$a^{-0.12+0.1\beta+0.71(\gamma-\vartheta)} R_0^{-0.67-0.49\beta-0.67\gamma+1.67\vartheta} \wp^{\frac{1}{14}[3\beta-5(\vartheta-\gamma-1)]}$$

The two scaling laws (81, 82) are general, since they do not depend on the physics behind the computation of $\Delta(Q,P)$, which determines the exponents $\beta, \gamma$. In (81), (82) the maximum dependence of the exponents is on the parameter $\vartheta$: this highlights the important role of the neoclassical poloidal flow damping effect in determining the form of the scaling law.

Let us now provide an estimate of $\beta, \gamma$. According to figure 6 the function $\Upsilon(Q,P)$ has a non-monotonic dependence on $Q$, weakly depending on $P$, with a maximum at $Q = Q_m(P) \sim 1.5$. We can take this maximum as a break between two $Q$ intervals. Roughly speaking, we approximate $\Upsilon(Q,P) \propto Q$ for $Q < Q_m$, and $\Upsilon(Q,P) \propto Q^{-1}$ for $Q > Q_m$. Of course, this approximation breaks downs near the maximum $Q_m$. Then, we expect $\mathcal{T}(Q_0, Q, P) \approx Q(Q_0 - Q)$ for $Q_0 < Q_m$, so that $m(Q_0, P) \propto Q_0^2/4$, and $\beta \sim 1$. Likewise, we expect $\mathcal{T}(Q_0, Q, P) \approx Q_0/Q - 1$ for $Q_0 > Q_m$, so that and $m(Q_0, P) \propto Q_0/Q_m - 1$, and $\beta \sim 0$. In both case there should be a weak $P$ dependence, namely $|\gamma| \ll 1$. These approximations are confirmed by the multivariate regression $m(Q_0, P) \propto Q_0^{1+\beta} P^\gamma$ shown in figure 7 for the two separate intervals $Q_0 < 1.5$, $Q_0 > 1.5$. Of course, the power-law representation of $m$ is justified by the relatively weak dependence of the scaling laws (81), (82) on $\beta, \gamma$.

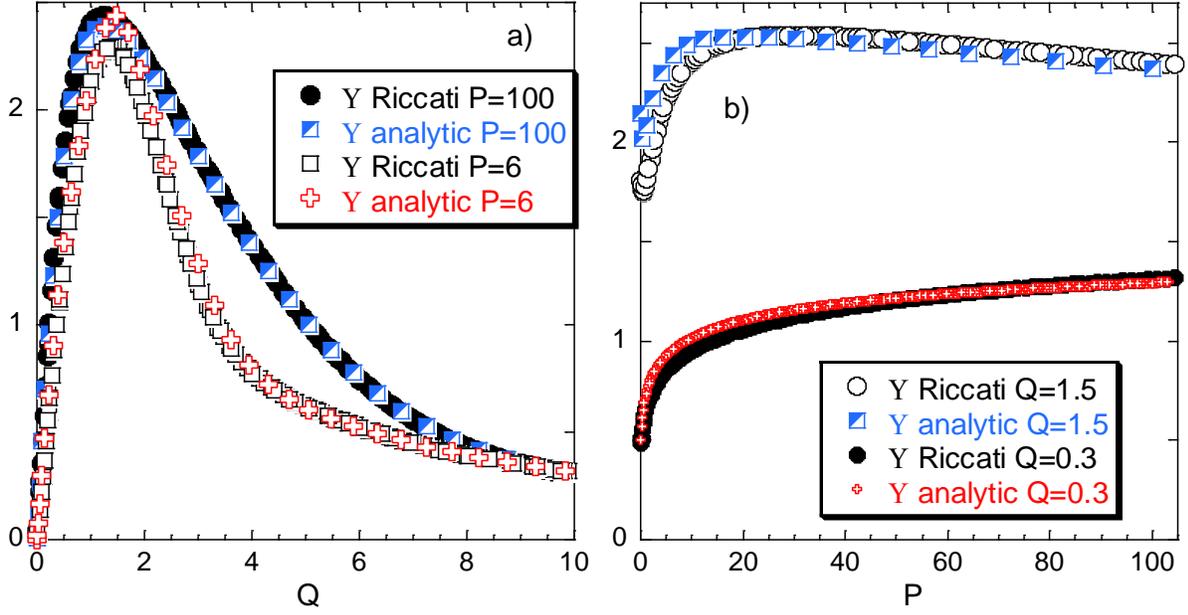

**Fig.6** $\Upsilon(Q,P)$ as function of $Q$ (a) and $P$ (b). Both the Fourier-Riccati computation, and the analytic estimate based on (68) are shown.

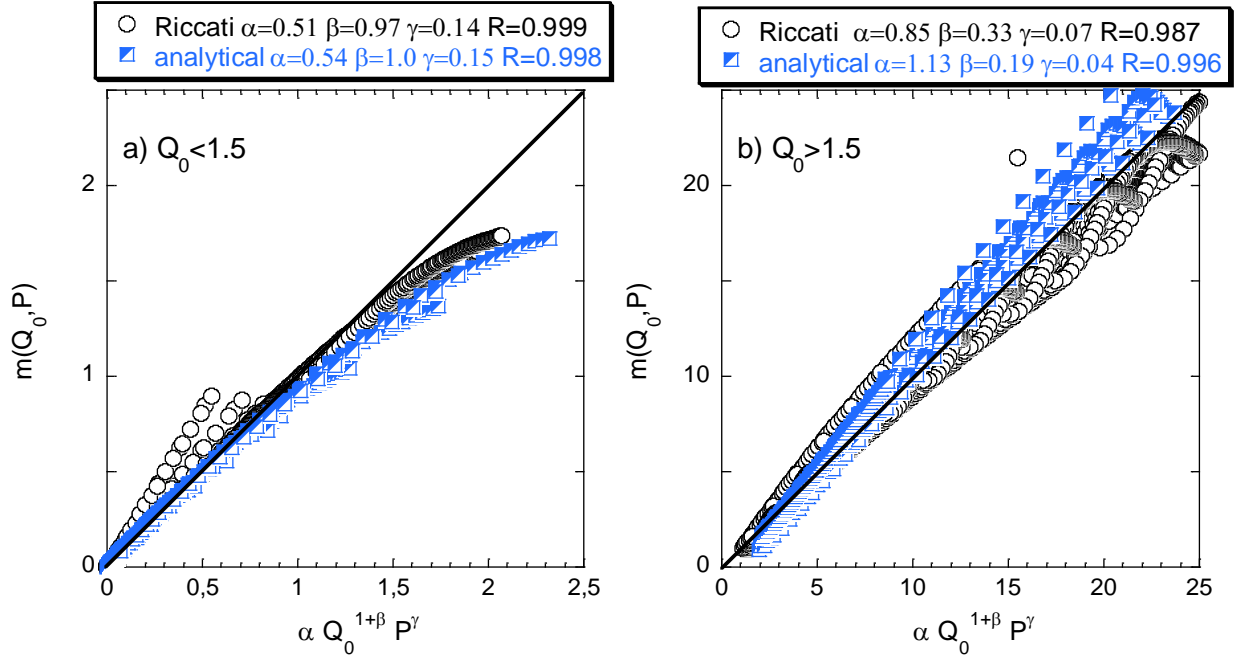

**Fig. 7** Multivariate regression $m(Q_0, P) = \alpha Q_0^{1+\beta} P^\gamma$ in two regions of the $Q$ parameter. The fit is made both on data from the Fourier-Riccati numerical method (black open circles), and on data from the analytical method (blue squares). The values obtained for $\alpha, \beta, \gamma$ are reported in the legends, as well as the Pearson correlation coefficient $R$. The straight black line is the bisector.

Hereafter, we take the $\beta, \gamma$ coming from the numerical $\Delta(Q, P)$. Since $\gamma$ is small, and $\beta$ is multiplied by small factors in the exponents, the variations of (81, 82) in the two cases $Q_0 \lessgtr 1.5$ are minimal. Let us take $\vartheta = 0.4$, according to our previous estimate. We get:

83) $\left(\frac{|b_r(b)|}{B_0}\right)_{thr,Alc} \propto n_e^{0.41} B_0^{-1.11} a^{-0.27} R_0^{-0.34}, \quad \vartheta = 0.4, \quad Q_0 < 1.5$

84) $\left(\frac{|b_r(b)|}{B_0}\right)_{thr,Alc} \propto n_e^{0.34} B_0^{-1.01} a^{-0.31} R_0^{-0.27}, \quad \vartheta = 0.4, \quad Q_0 > 1.5$

85) $\left(\frac{|b_r(b)|}{B_0}\right)_{thr,89P} \propto n_e^{0.22} B_0^{-0.90} a^{-0.21} R_0^{-0.58} \wp^{0.47}, \quad \vartheta = 0.4, \quad Q_0 < 1.5$

86) $\left(\frac{|b_r(b)|}{B_0}\right)_{thr,89P} \propto n_e^{0.38} B_0^{-1.06} a^{-0.33} R_0^{-0.21} \wp^{0.31}, \quad \vartheta = 0.4, \quad Q_0 > 1.5$

We also report the scaling laws obtained with $\vartheta = 0$, i.e with only the toroidal flow contribution in (73). This corresponds to the limit case $\tau_\theta \to 0$, where the poloidal flow is completely suppressed, giving no contribution to the viscous torque:

87) $\left(\frac{|b_r(b)|}{B_0}\right)_{thr,Alc} \propto n_e^{0.01} B_0^{-0.87} a^{0.05} R_0^{-1.14}, \quad \vartheta = 0, \quad Q_0 < 1.5$

88) $\left(\frac{|b_r(b)|}{B_0}\right)_{thr,Alc} \propto n_e^{-0.06} B_0^{-0.77} a^{0.01} R_0^{-1.07}, \quad \vartheta = 0, \quad Q_0 > 1.5$

89) $\left(\frac{|b_r(b)|}{B_0}\right)_{thr,89P} \propto n_e^{-0.07} B_0^{-0.78} a^{0.08} R_0^{-1.24} \wp^{0.62}, \qquad \vartheta = 0, \quad Q_0 < 1.5$

90) $\left(\frac{|b_r(b)|}{B_0}\right)_{thr,89P} \propto n_e^{0.09} B_0^{-0.94} a^{-0.04} R_0^{-0.88} \wp^{0.45}, \qquad \vartheta = 0, \quad Q_0 > 1.5$

Finally, we report the case $\vartheta = 0.5$, i.e complete dominance of the poloidal flow contribution in (73):

91) $\left(\frac{|b_r(b)|}{B_0}\right)_{thr,Alc} \propto n_e^{0.51} B_0^{-1.17} a^{-0.35} R_0^{-0.14}, \qquad \vartheta = 0.5, \quad Q_0 < 1.5$

92) $\left(\frac{|b_r(b)|}{B_0}\right)_{thr,Alc} \propto n_e^{0.44} B_0^{-1.07} a^{-0.39} R_0^{-0.07}, \qquad \vartheta = 0.5, \quad Q_0 > 1.5$

93) $\left(\frac{|b_r(b)|}{B_0}\right)_{thr,89P} \propto n_e^{0.3} B_0^{-0.93} a^{-0.28} R_0^{-0.4} \wp^{0.44}, \qquad \vartheta = 0.5, \quad Q_0 < 1.5$

94) $\left(\frac{|b_r(b)|}{B_0}\right)_{thr,89P} \propto n_e^{0.46} B_0^{-1.09} a^{-0.40} R_0^{-0.04} \wp^{0.27}, \qquad \vartheta = 0.5, \quad Q_0 > 1.5$

Note a positive dependence on the additional/ohmic power fraction $\wp$. By comparing (87-90) with (91-94) we realize the crucial importance of the parameter $\vartheta$, as said before. These scaling laws are very close to those obtained in [6], where an analysis similar to ours, apart the use of the two-fluids drift-MHD to compute the $\Delta(Q, P)$, is presented. In doing the comparison with [6] we add the exponents of the minor and major radii of our expressions, since they both represent a dependence on the machine size. We find that (91-94) are nearly identical to the scaling laws (152), (153), (157), (158) of [6], which refer to the case of poloidal flow dominance over toroidal flow in the torque-balance equation ($\vartheta = 0.5$ in our analysis). Likewise, formulas (87-90) are almost identical to formulas (151), (154), (156), (159) obtained in [6] in the case where the toroidal flow dominates over the poloidal flow in the torque balance equation ($\vartheta = 0$ in our analysis). We draw the conclusion that, as far as the EF threshold is concerned, it is of little importance the use of single-fluid or two-fluids MHD in computing $\Delta(Q, P)$. In other words, the details of $\Delta(Q, P)$ little affect the EF threshold: the crucial element is the poloidal-flow damping term in the torque balance equation.

Finally, we discuss the comparison with the experimental results. A large database, covering both H-mode and L-mode tokamak plasma, is fitted by the following scaling law for the $n = 1$, EF threshold [20]:

$$95) \quad \left(\frac{|b_r(b)|}{B_0}\right)^{n=1}_{thr,exp} \propto n_e^{0.58\pm0.06} B_0^{-1.13\pm0.07} R_0^{0.1\pm0.07}$$

We remark that the quality of the fit is not very good (see figure 2 of [20]), hence this scaling law should be taken with some reserve. A better fit, but on a smaller database is presented in the same paper for $n = 2$ EF threshold [20]

$$96) \quad \left(\frac{|b_r(b)|}{B_0}\right)^{n=2}_{thr,exp} \propto n_e^{1.07\pm0.09} B_0^{-1.52\pm0.2} R_0^{1.46\pm0.09}$$

The two fits (95), (96) are remarkably different as far as density and machine dimensions are concerned. Expressions (83, 91, 92, 94) referring to $\vartheta = 0.4 \div 0.5$ are close to (95) as far as density and magnetic field exponents are concerned. There is a discrepancy concerning the machine size exponent, a general issue in this kind of theoretical models [6]. Instead, the expressions obtained with $\vartheta = 0$ are quite far from (95). However, the first case is much more likely according to our estimates.

## 8. Conclusions

In this work we re-examined the linear plasma response to a static EF in the single fluid visco-resistive MHD, by focusing on several basic aspects of the problem. We validated the radial Fourier transform method by comparison with a completely different numerical technique. Then we derived the new analytical Δ' general formula (68), which agrees fairly well with the numerical prediction. This formula describes the Δ' features much better than previous asymptotic regimes expressions [2]. On the basis of our Δ' estimate, the EF penetration threshold has been derived, and we pointed out the crucial role of the neoclassical poloidal flow damping term. In fact, only if the poloidal flow modification dominates over the toroidal flow modification in the torque-balance equation, the prediction is similar to the experimental scaling law. This is the same result as obtained in the most recent two-fluids analysis [6]. Since the scaling laws here derived are almost identical to those of [6], we draw the obvious conclusion that the choice between single-fluid or two-fluids MHD is not crucial in this specific problem. As far as the comparison with the experimental results is concerned, a discrepancy on the machine size dependence persists, a general issue in this kind of theories. On the basis of our analysis, we suggest seeking a refinement of the EF threshold prediction not in the Δ' computation, rather in the modelling of the poloidal flow damping effect, which at present is included in the torque-balance equation with a heuristic expression.

**Acknowledgment** The author wish to thank Italo Predebon, Fabio Sattin, Dominique F. Escande for helpful discussions.